\documentclass[3p,times,procedia]{elsarticle}
\usepackage{nupha_ecrc}

\volume{00}

\firstpage{1}

\journalname{Nuclear Physics A}

\runauth{}

\jid{nupha}

\jnltitlelogo{Nuclear Physics A}

\usepackage{amssymb}

\usepackage[figuresright]{rotating}

\begin{document}

\begin{frontmatter}



\dochead{XXVIth International Conference on Ultrarelativistic Nucleus-Nucleus Collisions\\ (Quark Matter 2017)}

\title{Measurements of $\Lambda_c^+$ and $D_s^+$ productions in Au+Au collisions at $\sqrt{s_{\rm NN}}$ = 200 GeV from STAR}


\author{Long Zhou, for the STAR Collaboration}

\address{No. 96, Jinzai road, University of Science and Technology of China, Hefei, China}

\begin{abstract}
We report the first measurement of $\Lambda_c^+$ (average of $\Lambda_c^+$ and $\bar
{\Lambda}_c^-$) baryon production and significantly improved measurement of $D_s^+$ (average of $D_s^+$ and $D_s^-$) meson production, within the rapidity range of $\left|y\right| < 1$  in Au+Au collisions at $\sqrt{s_{\rm NN}}$ = 200 GeV from the STAR experiment. The transverse momentum spectra of $\Lambda_c^+$ and $D_s^+$ as well as their yield ratios to that of $D^0$ meson are also presented and compared to theoretical calculations.
Both the $\Lambda_c^+/D^0$ and $D_s^+/D^0$ ratios are found to be significantly enhanced in Au+Au collisions with respect to those in p+p collisions predicted by PYTHIA. A model calculation including coalescence hadronization and thermalized charm quarks is consistent with the measured $\Lambda_c^+/D^0$ ratio.  On the other hand, the TAMU model seems to underestimate the $D_s^+/D^0$ ratio measured in Au+Au collisions in the applicable kinematic region.

\end{abstract}


\end{frontmatter}


\section{Introduction}
\label{intro}
Quark coalescence has been proposed as a candidate to explain the Number-of-Constituent-Quark scaling for meson/baryon elliptic flow as well as the enhancement in the baryon-to-meson ratios observed in heavy-ion collisions in the intermediate transverse momentum range ($2<p_{T}<6$ GeV/c) for both light and strange flavor hadrons~\cite{coale_2,coale_1}. If the coalescence mechanism also plays a significant role for charm quark hadronization inside the hot and dense medium, one would expect enhancements in the charm-strange meson and charm baryon yields in heavy-ion collisions~\cite{ds_prl,ko,shm,greco}. The magnitudes of the enhancements are sensitive to the QGP dynamics, e.g. the degree of thermalization for charm quarks, the amount of strangeness enhancement, etc. Knowledge of the yields for different charm hadrons is also critical for determining the total charm quark yield in heavy-ion collisions.

The Heavy Flavor Tracker (HFT), installed at the STAR experiment between 2014 and 2016, was designed to extend STAR's capability of measuring open charm hadrons via direct reconstruction of displaced decay vertices. The HFT consists of 4 layers of silicon detectors. The outer layer is the Silicon Strip Detector (SSD). The Intermediate Silicon Tracker (IST), consisting of one layer of silicon pad sensors, is located inside the SSD. Two layers of the Silicon Pixel Detector (PXL) are inside the IST. The PXL detector provides excellent hit position resolution for precise measurements of displaced vertices. The track pointing resolution of the HFT exceeds the design goal of 55 ${\mu}$m for kaons with $p_{T} = 750$ MeV/$c$~\cite{contin}.
\section{Data Analysis}
\begin{figure}
	\centering	
	\includegraphics[width=0.48\linewidth]{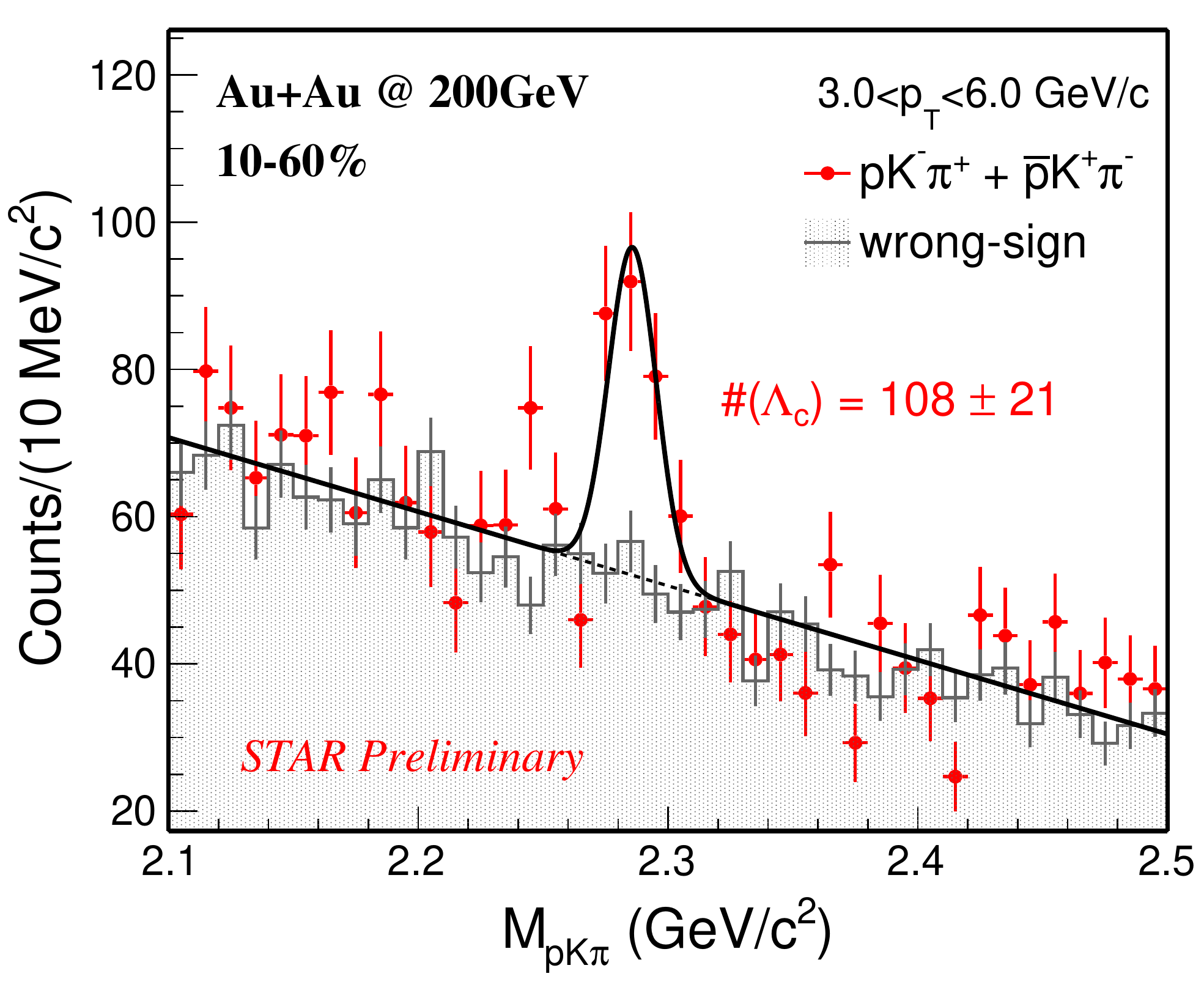}	
	\includegraphics[width=0.48\linewidth]{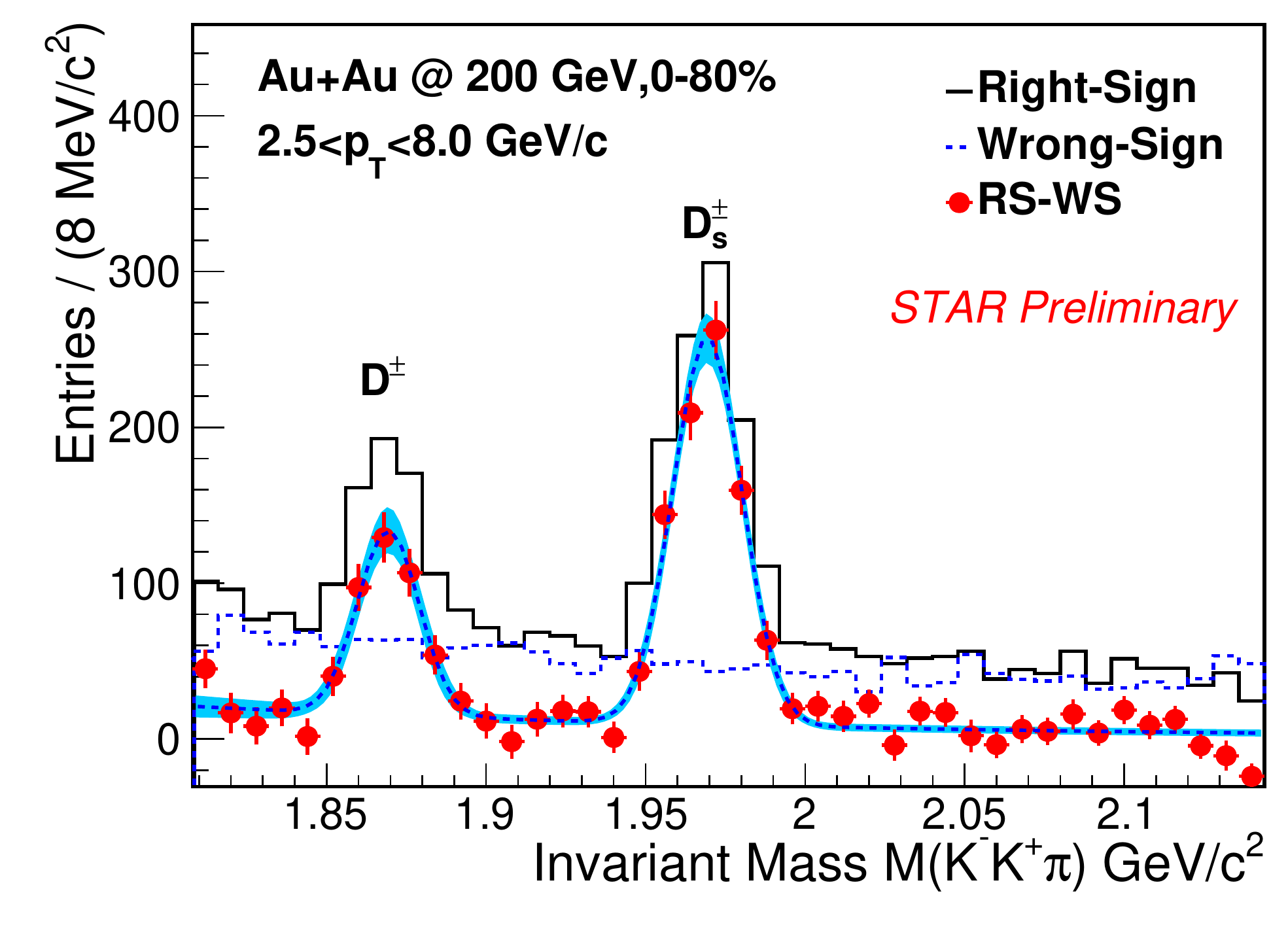}	
	\caption{(Color online) The invariant mass distribution for $\Lambda_c^+$ (left panel) and $D_s^+$ (right panel) candidates in 10-60\% and 0-80\% central Au+Au collisions at $\sqrt{s_{\rm NN}}$ = 200 GeV, respectively. A first-order polynomial function is used to describe the combinatorial background and a Gaussian function for the signal. The cyan band in the right panel is the 68\% confidence interval for the fit function.  }
	\label{fig:ds_lc_sig}
\end{figure}

About 900 million minimum-bias Au+Au events at $\sqrt{s_{\rm NN}} = 200$ GeV are used for the results presented here. The reconstructed primary vertices are required to be less than 6 cm away from the center of the STAR detector along the beam direction to ensure uniform HFT acceptance. Charged tracks are reconstructed in the Time Projection Chamber (TPC) and are required to have at least one hit on each layer of the IST and PXL. The particle identification of daughter tracks at mid-rapidity ($\left|y\right|$$<$1) is carried out using the energy loss ($dE/dx$ ) measured by the TPC and 1/$\beta$ ($\beta=v/c$ , where $v$ is the particle velocity and $c$ is the speed of light) by the Time-of-Flight (TOF) detector.  For each track, the $dE/dx$ is required to be within 3 and 2 standard deviations from the expected values for pions and kaons, respectively. If a track is matched to a TOF hit, a cut of $\left|1/\beta - 1\right|$ $<$ 0.03 is applied. The $\Lambda_c^+$ (sum of $\Lambda_c^+$ and $\bar{\Lambda}_c^-$) baryon is reconstructed through the decay channel:  $\Lambda_{c}^{+}{\rightarrow}p+K^{-}+\pi^{+}$ with a branching ratio of 6.35 $\pm$ 0.33 \% , and the $D_s^+$ meson (sum of $D_s^+$ and $D_s^-$) is reconstructed through the hadronic decay channel: $D_s^+\rightarrow\phi(1020)+\pi^+\rightarrow K^{+}+K^{-}+\pi^+$, whose branching ratio is $2.27 \pm 0.08$ \% \cite{pdg_2016}. Geometric and kinematic cuts are applied to reduce the combinatorial background. The left and right panels of Fig.~\ref{fig:ds_lc_sig} show the reconstructed $\Lambda_c^+$ signal in 10-60\%  and $D_s^+$ signal in 0-80\% centrality class, respectively.  A clear $D^+$ (sum of $D^+$ and $D^-$) signal is also observed in the same decay channel as $D_s^+$ mesons with a branching ratio of $0.277 \pm 0.10$ \% \cite{pdg_2016}.

\begin{figure}
	\centering	
	\includegraphics[width=0.48\linewidth]{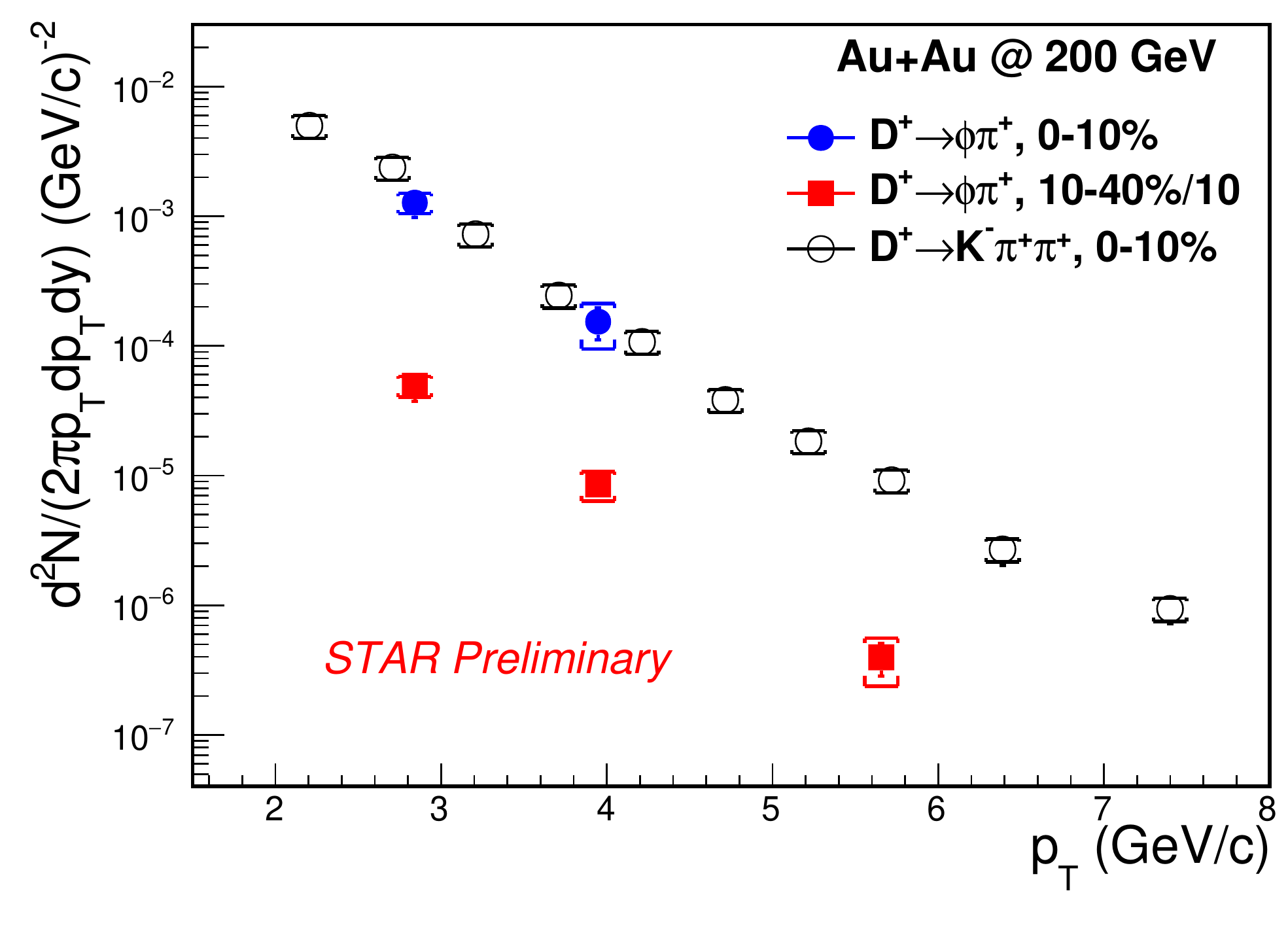}	
	\includegraphics[width=0.48\linewidth]{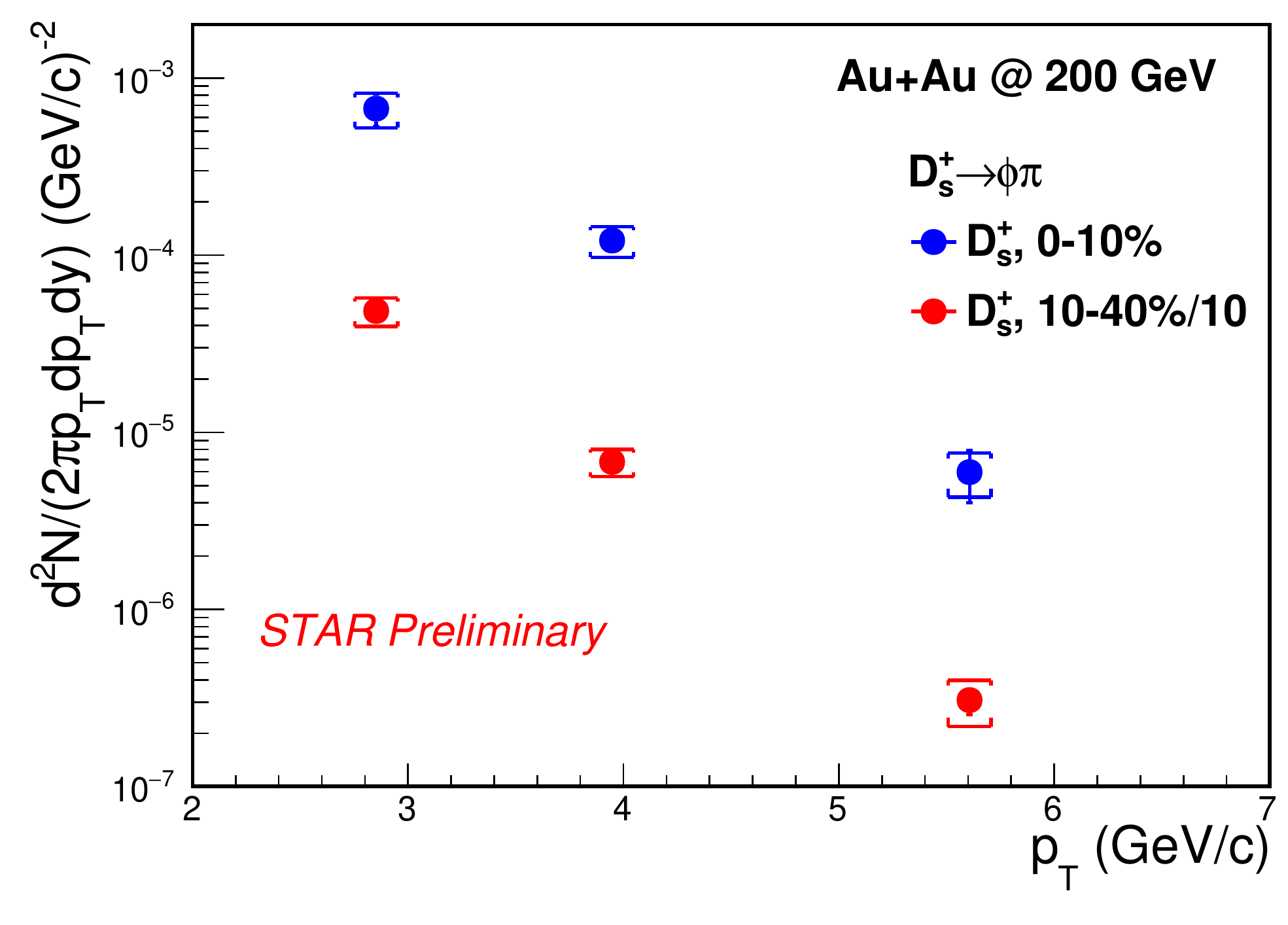}	
	\caption{(Color online) The invariant yield distributions for $D^+$ (left) and $D_s^+$ (right) mesons in 0-10\% and 10-40\% central Au+Au collisions at $\sqrt{s_{\rm NN}}$ = 200 GeV. The blue and red markers in both left and right panels represent the 0-10\% and 10-40\% centrality classes, respectively. The black  open circles in the left panel  show the $D^+$ invariant yield measured via $D^{+}\rightarrow K^{-}+\pi^{+}+\pi^{+}$. }
	\label{fig:ds_dpm_pt_spectra}
\end{figure}

The invariant yields of $D^+$ (average of $D^+$ and $D^-$)  and $D_s^+$ (average of $D_s^{+}$ and $D_s^{-}$) measured in 0-10\% and 10-40\% central Au+Au collisions are presented in the left and right panels of Fig.~\ref{fig:ds_dpm_pt_spectra}, respectively. In the left panel, the invariant yields of $D^{+}$ in 0-10\% centrality class is also measured through another decay channel, i.e. $D^{+}\rightarrow K^{-}+\pi^{+}+\pi^{+}$ (branching ratio 9.46 $\pm$ 0.24 \%)~\cite{star_dpm} which is shown as the black open circles.  Even though the branch ratios for those two decay channels differ by about a factor of 35, the corrected invariant yields show very good agreement.

\begin{figure}
	\centering	
	\includegraphics[width=0.48\linewidth]{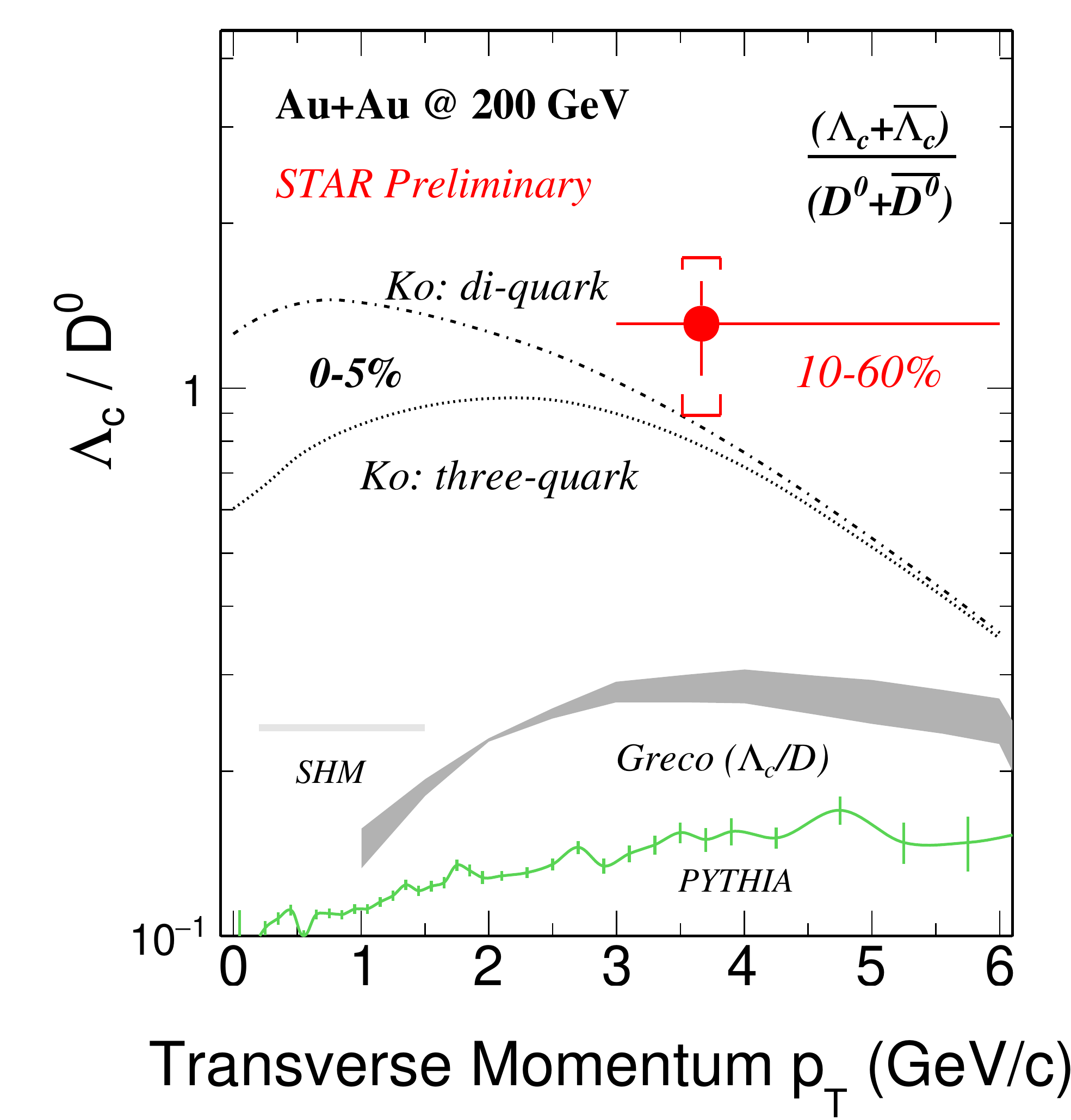}	
	\includegraphics[width=0.48\linewidth]{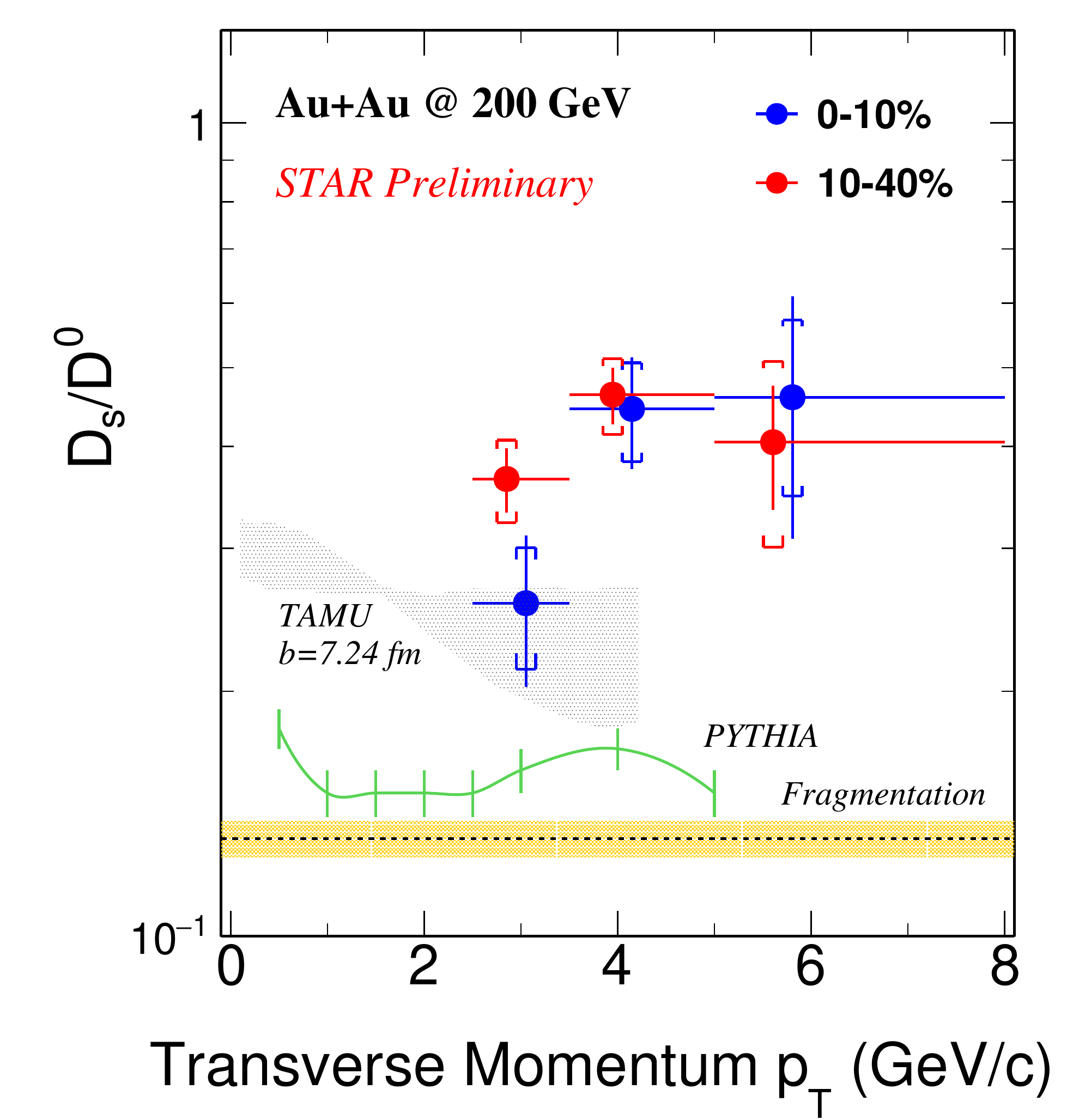}	
	\caption{(Color online) The $\Lambda_c^+/D^0$ (left) and $D_s^+/D^0$ (right) yield ratios as a function of $p_T$ in Au+Au collisions at $\sqrt{s_{\rm NN}}$ = 200 GeV.  Different model predictions for both $\Lambda_c^+/D^0$ and $D_s^+/D^0$ yield ratios are also presented for comparison.}
	\label{fig:yields_ratio}
\end{figure}

The $\Lambda_c^+$ and $D_s^+$ yields are compared to the $D^0$ yield~\cite{stard0} in Au+Au collisions to study the modifications to their production in the QGP due to the potential involvement of charm quarks in the coalescence hadronization and the strangeness enhancement. The left panel of Fig.~\ref{fig:yields_ratio} shows the comparison of the  measured $\Lambda_c^+/D^0$ yield ratio as a function of $p_T$ in 10-60\% centrality class to several model calculations~\cite{ko,greco,shm} with different degrees of thermalization for charm quarks in the medium and different implementations of the coalescence mechanisms. The Ko model calculations with full charm quark thermalization and either di-quark or three-quark coalescence mechanism are consistent with data.
The $D_s^+/D^{0}$ yield ratios as function of $p_T$ in 0-10\% and 10-40$\%$ central in  Au+Au collisions at $\sqrt{s_{\rm NN}}$ = 200 GeV are shown in the right panel of Fig.~\ref{fig:yields_ratio}, and  compared with the world-data average of the charm hadron fragmentation ratio (orange band)~\cite{frac_avg},  the PYTHIA prediction (green line)~\cite{pythia} as well as the TAMU model calculation (gray band).  The PYTHIA prediction of  the $D_s^+/D^0$ ratio is consistent with the world-data average of the fragmentation ratio,  and both of them are significantly below the measured values in central and semi-central Au+Au collisions.  On the other hand, the TAMU model~\cite{ds_prl} predicts an enhanced $D_s^+/D^0$ ratio for $p_T<$ 4 GeV/c as a consequence of the charm quark coalescence with thermalized strange quarks in the QGP, which seems to underestimate the measured enhancement in the corresponding $p_T$ range. Furthermore, the TAMU model calculation is expected to be consistent with the world-data average at high $p_T$ since fragamentation is believed to be the dominant hadronization mechanism in this kinematic range. Therefore, other mechanisms might be needed to explain the observed significant enhancement at high $p_T$.

\section{Summary}
We present the first measurement of $\Lambda_c^+$ production in Au+Au collisions at $\sqrt{s_{\rm NN}}$ = 200 GeV using the HFT at the STAR experiment. The $\Lambda_c^+/D^{0}$ yield ratio is 1.3 $\pm$ 0.3(stat) $\pm$ 0.4(sys) for 3 $< p_T <$ 6 GeV/c in 10-60\% centrality class, which is significantly larger than the PYTHIA prediction for p+p collisions. The measured $\Lambda_c^+/D^0$ yield ratio can be described by the Ko model calculations with both di-quark and three-quark coalescence mechanisms and full charm quark thermalization. The latter is consistent with the measured $D^0$ elliptic flow~\cite{star_d0_v2}.
 The $D_s^+/D^0$ ratio is higher than that in p+p collisions predicted by PYTHIA at the intermediate $p_{T}$ range,  indicating an enhancement of $D_s^+$ production in Au+Au collisions. The TAMU model seems to underestimate the observed enhancement below 4 GeV/c, while the enhancement seen in data at higher $p_T$ remains a challenge for the model.

\section*{Acknowledgement}
This work was supported in part by the Major State Basic Research Development Program in China with grant no. 2014CB845402 and the National Natural Science Foundation of China with grant no. 11375184.







\end{document}